\documentclass[useAMS,usenatbib,letterpaper]{mn2e}
\usepackage[totalwidth=480pt,totalheight=680pt,layoutvoffset=0.5cm]{geometry}
\usepackage{graphicx}
\usepackage{amsmath}
\usepackage{astrobib_mnras2e}
\usepackage{times}
\usepackage{fixltx2e}

\newcommand{\cff}{\chi_{f,0}}

\title{Optimal Redshift Weighting For Baryon Acoustic Oscillations}
\author[Zhu et al]{Fangzhou Zhu$^1$, 
Nikhil Padmanabhan$^1$, 
Martin White$^2$ \\
\scriptsize $^1$ Dept. of Physics, Yale University, New Haven, CT 06511 \\
\scriptsize $^2$ Dept. of Physics and Astronomy, U.C. Berkeley, Berkeley, CA \\
}

\begin{document}

\maketitle
\begin{abstract}
Future baryon acoustic oscillation (BAO) surveys will survey very large
volumes, covering wide ranges in redshift. 
We derive a set of redshift weights to compress the information in the
redshift direction to a small number of modes. We suggest that such 
a compression preserves almost
all of the signal for most cosmologies, while giving high signal-to-noise
measurements for each combination.  We present some toy models and simple
worked examples. As an intermediate step, we give a precise meaning to the
``effective redshift'' of a BAO measurement.
\end{abstract}

\section{Introduction}
\label{sec:intro}

The baryon acoustic oscillation (BAO) method has now become an essential
component of the dark energy toolbox by mapping the expansion history of
the Universe \citep[see][for recent reviews]{Wei13,MorWeiWhi14}.
Current sets of surveys \citep{Blake12, AndersonDR11} have achieved 
distance accuracies at the per cent level when averaged across relatively
wide redshift bins; the next generation of surveys \citep{Sumire,DESI,Euclid,WFIRST} 
aim to improve both the redshift resolution and the redshift range of these
measurements.
This paper aims to develop techniques to robustly map the distance-redshift
relation for these surveys.

Conceptually, one can improve the resolution of the reconstructed
distance-redshift relation by splitting samples into narrower redshift
bins and repeating the traditional analysis on these.
However, such a procedure has a significant disadvantage for BAO measurements:
the robustness of the BAO technique derives from focusing only on the BAO
feature in the correlation function or power spectrum, marginalizing over the broadband shape.
A reliable measurement of distances therefore requires a significant detection
of the BAO feature in the correlation function and low signal-to-noise
detections can lead to very non-Gaussian likelihood functions for the inferred
distance \citep[eg.][]{Blake12, Xu2013}
Splitting samples into multiple bins exacerbates this problem.
Binning also has other disadvantages: the choice of bins is often arbitrary
and splitting a sample into completely disjoint bins loses signal across bin
boundaries unless a large number of cross-spectra are included.  While some
of these disadvantages may be overcome, they can add significantly to the
complexity of the analysis.

Rather than breaking the sample into multiple redshift slices, we instead
seek a compression of the information in the redshift direction onto a
small number of `modes', or weighted integrals.  This allows us to work
with relatively wide redshift bins while retaining sensitivity to evolution
of the distance-redshift relation and gives a firm meaning to the `effective
redshift' of the sample.
Our approach is reminiscent of principal component analysis or optimal
subspace filtering, though being technically somewhat different.

To do so, we start by parametrizing the distance redshift relation relative
to a fiducial choice of cosmology (see \citet{Percival07} for related 
efforts to fit a distance redshift relation using BAO measurements). We demonstrate that simple parametrizations
capture a broad range of possible distance redshift relations over wide 
redshift intervals. 
Given a particular choice of parametrization, we 
then derive appropriate weighted integrals (in redshift) of the correlation
function that are (under particular assumptions) optimal  estimates of the parameters
in the distance-redshift relation. These integrals may be thought of as
generalizations of simple redshift bins, designed to efficiently combine the 
information at different redshifts. 

While we argue in this paper for a model-agnostic parametrization,
different choices may be more appropriate for more specific applications.
For instance, one may choose the equation of state of dark energy to
parametrize the distance-redshift relation.
The modes would then be different from the ones we focus on for much of
the presentation, but the overall approach would remain unchanged.  We
give a simple example later to illustrate this flexibility.

The outline of this paper is as follows.  We introduce our parametrization
of the distance-redshift relation in section \ref{sec:dz}, showing that it
fits a wide range of cosmologies to the sub per cent level.
Section \ref{sec:weights} introduces the weighted correlation functions
and our optimal weights, including a toy model to demonstrate the improvements
that can be expected by weighting.
Section \ref{sec:bao} specializes our discussion to the case of BAO and
describes a possible implementation of the method.  We conclude in
section \ref{sec:discuss} with a discussion of our results and directions
for future development.  Some technical details are relegated to an appendix.

\section{Parametrizing the Distance Redshift Relation}
\label{sec:dz}

We begin by parametrizing the distance-redshift relation
as a small fluctuation
about a fiducial relation.  This scheme is standard in BAO studies, as one
typically uses a fiducial cosmology to convert the angular positions and
redshifts of the galaxies into 3D positions before measuring the clustering
signal.  Motivated by inflation and supported by recent observations
\citep{PlanckXVI,AndersonDR11} we shall assume throughout that the spatial
hypersurfaces are flat, and we shall denote the comoving radial distance as
$\chi$. 

We choose a pivot redshift $z_0$ within the redshift range of a planned survey and denote the fiducial comoving 
distance at $z_0$ by $\chi_{f,0}$. We defer the reader to the end of this section for a discussion about the 
choice of $z_0$, while merely pointing out the analysis below doesn't depend on a particular choice of $z_0$. 
We model the true distance-redshift relation $\chi(z)$ as 
\begin{equation}
  \frac{\chi(z)}{\chi_{f}(z)} = \alpha_0
  \left(1 + \alpha_1 x + \frac{1}{2} \alpha_2 x^2 \right) 
\end{equation}
where $1 + x \equiv \chi_{f}(z)/\chi_{f,0}$.
The parameters $\alpha_0$, $\alpha_1$ and $\alpha_2$ are specified by matching the distance-redshift relation
and its derivatives at $z_0$ :
\begin{align}
	\alpha_0 &= \frac{\chi_0}{\cff} \\
	\alpha_1 &= \frac{H_{f,0} \cff}{H_0 \chi_0} - 1 \\
	\alpha_2 &= (1+\alpha_1)\chi_{f,0}\left[H'_{f,0} - \alpha_0(1+\alpha_1)H'_0\right] - 2\alpha_1
\end{align}
where $H'_{f,0} = H'_f(z_0)$ and $H'_0 = H'(z_0)$. An explicit calculation of $\alpha_1$ and $\alpha_2$ is in the appendix.

\begin{figure}
\includegraphics[width=0.5\textwidth]{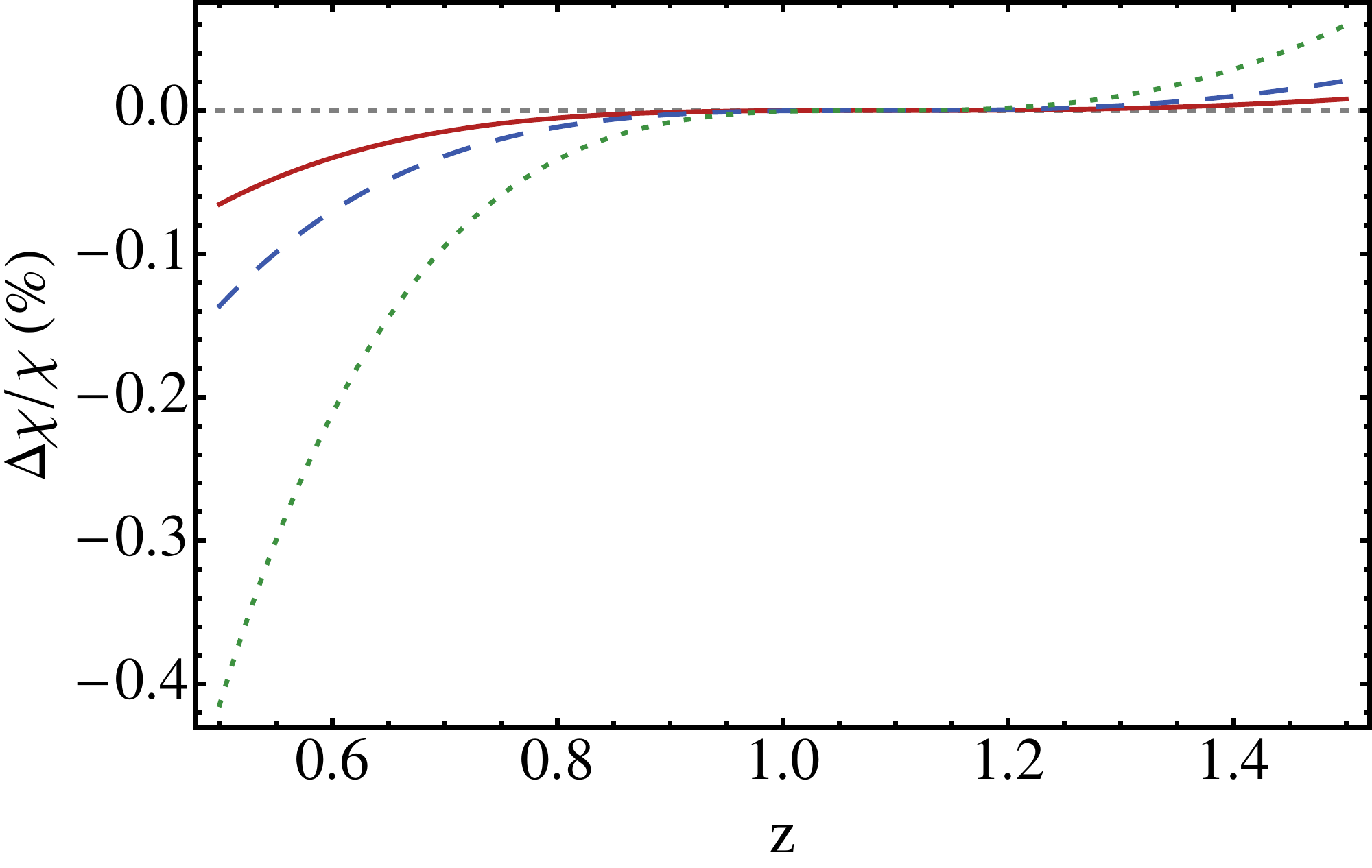}
\caption{The fractional error in our proposed distance parametrization for three
example cosmologies. The fiducial
cosmology here is a flat $\Lambda$CDM cosmology with $\Omega_M=0.274$ and $h=0.7$. The 
solid [red] line is an $\Omega_M=0.2$, $h=0.7$ cosmology; the short dashed line is our
fiducial cosmology with a dark energy equation of state $w=-0.7$, and the dotted
[green] line corresponds to $\Omega_M=1$ and $h=1$. The pivot redshift, $z_0 = 1.08$
is the effective redshift of the redshift range between $z=0.5$ and $1.5$, similar to the 
the redshift coverage of planned redshift surveys like DESI and Euclid.}
\label{fig:errR}
\end{figure}

\begin{figure}
\includegraphics[width=0.5\textwidth]{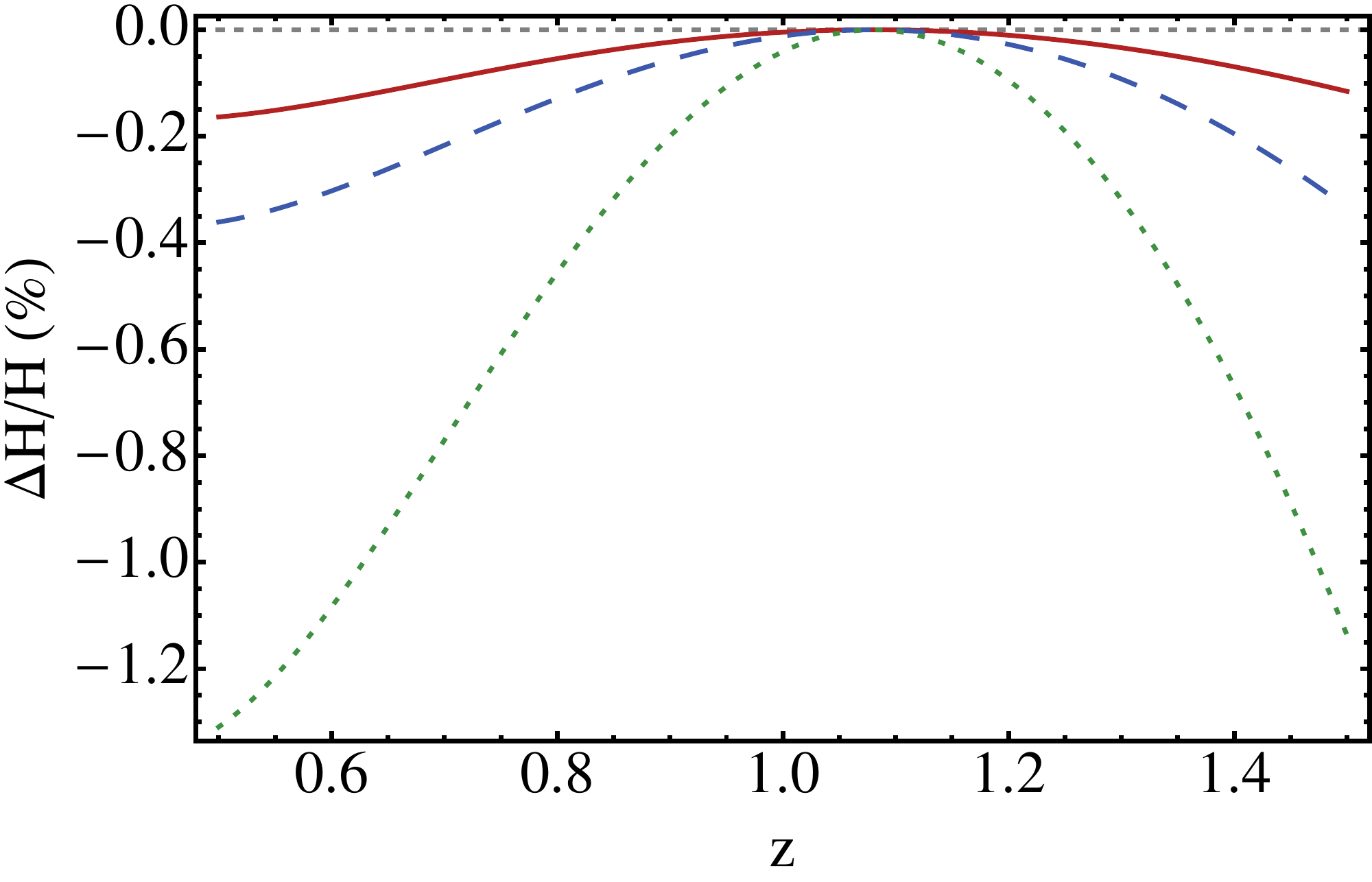}
\caption{The same as Fig.~\ref{fig:errR} except now showing the error in the Hubble parameter $H(z)=1/\chi'(z)$. The errors in the plot
are larger, reflecting the fact that our approximation is first order for $H$, albeit being second order for $\chi$.}
\label{fig:errH}
\end{figure}

\begin{figure}
\includegraphics[width=0.5\textwidth]{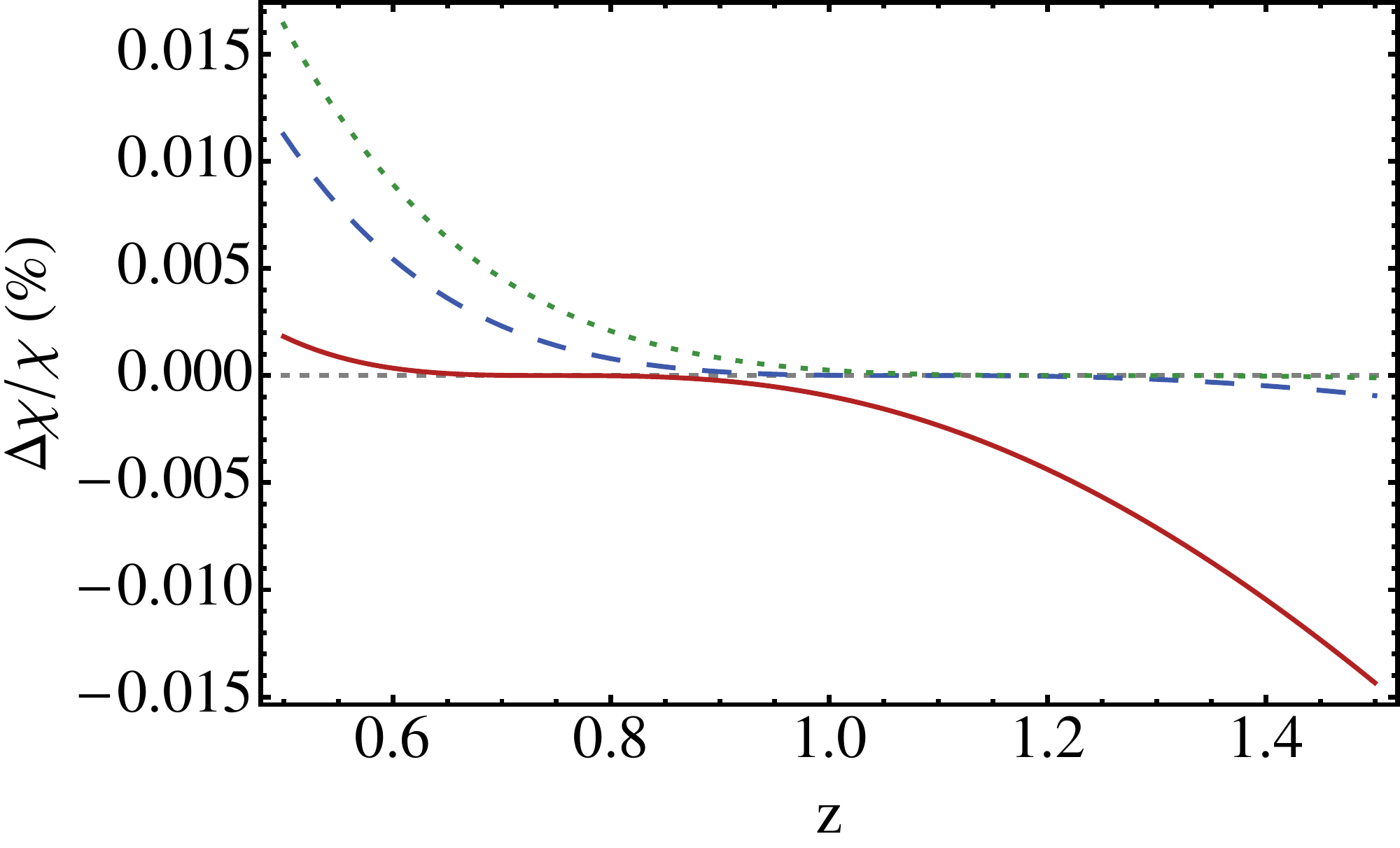}
\caption{The impact of changing $z_0$ on the distance redshift relation. The fiducial cosmology here is the same as the 
previous two figures, while the true cosmology has $\Omega_M=0.3$, $h=0.7$ and $w=-1$. The dashed [blue] line assumes
$z_0=1.08$ (equal to the effective redshift), the solid [red] line corresponds to $z_0=0.75$, while the dotted [green] 
line assumes $z_0=1.25$. The reconstructed distance-redshift relations all agree to better than 0.02 per cent over the entire 
redshift range.}
\label{fig:errz0}
\end{figure}

This parametrization to the second order can recover the distance-redshift
relation to sub-percent levels across a wide range of redshifts and
cosmologies.
Fig.~\ref{fig:errR} displays the fractional error in the
parametrization between $z=0.5$ and $1.5$ for three example cosmologies. 
These span a wide range of cosmological parameters and are not meant
to represent currently viable models.
Even for the extreme $\Omega_M=1$ case,
the error is less than $0.5\%$ over the entire redshift range. 

Fig.~\ref{fig:errH} shows a similar plot of the fractional error in the Hubble
parameter $H(z) = 1/\chi'(z)$. A second order approximation
in distance translates to a first order one in $H$ and explains the larger errors.
Despite that, for cosmologies close to the fiducial model,
the accuracy is still at the sub-percent level. 

It is worth noting at this point that $z_0$ is a meta-parameter which can
be chosen for convenience, rather than a property of the sample under
consideration.  For best results the fiducial redshift should be chosen close
to the center of the redshift range, for example at the `effective' or
pair-weighted redshift of the sample. Fig.~\ref{fig:errz0} highlights this 
freedom in the choice of $z_0$, demonstrating that the reconstructed distance-redshift
relations in each of these cases agree to better than 0.02 per cent, much smaller than
the expected statistical uncertainty in these measurements.

\section{Optimal Weights}
\label{sec:weights}

\subsection{Derivation}
\label{sec:optderive}


We start by deriving the general formalism for optimal linear weights.
The derivation is modeled on \cite{TTH}; we summarize it here for completeness,
while specializing to the particular problem at hand. 
Imagine measuring the correlation function in $n$ $z$-bins.
The uncompressed correlation function data vector is
\begin{equation}
\boldsymbol{\Xi}(r)=\left( \begin{array}{c}
\xi(r,z_1) \\
\xi(r,z_2)\\
\vdots\\
\xi(r,z_n)\end{array} \right)
\end{equation} 
Weighting the redshift slices by the $n$-dimensional vector
$\boldsymbol{w}^t=\left(w_1,w_2,\cdots, w_n\right)$,  we linearly compress the
data into one single $z$-bin according to these weights.
\begin{equation}
\xi_w(r)=\boldsymbol{w}^t\boldsymbol{\Xi}(r)
\end{equation}
where $\xi_w$ is the weighted correlation function. The mean and covariance $\boldsymbol{C}_w$ of the compressed data are
\begin{align}
\langle\xi_w(r)\rangle &=\boldsymbol{w}^t\langle \boldsymbol{\Xi}(r)\rangle \\
\langle \xi_w(r)\xi_w^t(r)\rangle - \langle \xi_w(r)\rangle \langle \xi_w^t(r) \rangle &= \boldsymbol{w}^t\boldsymbol{C}\boldsymbol{w}
\end{align}
where $\boldsymbol{C}$ is an $n\times n$ matrix and $\boldsymbol{C}_w = \boldsymbol{w}^t \boldsymbol{C}\boldsymbol{w}$ is a scalar.
We write our model correlation functions at distance $r$ in different $z$-bins
$\boldsymbol{M}(r)$
in the same format as $\boldsymbol{\Xi}$. Using the
redshift weights $\boldsymbol{w}$, we linearly compress the model correlation
function data in the same fashion, 
\begin{equation}
m(r)=\boldsymbol{w}^t \boldsymbol{M}(r)
\end{equation}
The model correlation function depends on parameters
$\boldsymbol{\Theta} = \left\{\theta_1, \theta_2, \cdots \right\}$. 
Restricting to a single unknown parameter, the error on $\theta_i$ is
$\left(\mathcal{F}_{ii}\right)^{-1/2}$,
where $\mathcal{F}_{ii}$ is the $i$-th diagonal element of the Fisher
information matrix. This is given by
\begin{equation}
\mathcal{F}_{ii} = m_{,i}^t \boldsymbol{C}_w^{-1} m_{,i}
\end{equation}
where $m_{,i}$ is the derivative of the model correlation function $m$ with respect to parameter $\theta_i$. Since $\boldsymbol{C}_w$ is a scalar, 
\begin{equation}
\mathcal{F}_{ii}=\frac{m_{,i}^t m_{,i}}{\boldsymbol{w}^t \boldsymbol{C} \boldsymbol{w}}
=\frac{\boldsymbol{w}^t \boldsymbol{M}_{ii} \boldsymbol{w}}{\boldsymbol{w}^t \boldsymbol{C} \boldsymbol{w}}
\end{equation}
where $\boldsymbol{M}_{ij}=\boldsymbol{M}_{,i}\boldsymbol{M}_{,j}^t$ is
an $n\times n$ matrix.
Our goal is to find the weights that minimize the error of $\theta_i$, and therefore maximize $\mathcal{F}_{ii}$.

Since $\mathcal{F}_{ii}$ is unchanged under a rescaling of $\boldsymbol{w}$,
we constrain $\boldsymbol{w}^t
\boldsymbol{C}\boldsymbol{w} = 1$ for convenience. 
Introducing a Lagrange multiplier $\lambda$, we maximize
\begin{equation}
  \mathcal{G}=\boldsymbol{w}^t \boldsymbol{M}_{ii}\boldsymbol{w}
  -\lambda\left(\boldsymbol{w}^t \boldsymbol{C}\boldsymbol{w} - 1 \right).
\end{equation}
Setting the gradient of $\mathcal{G}$ with respect to $\boldsymbol{w}$
to zero yields
\begin{equation}\label{eq:grad}
  \boldsymbol{M}_{ii}\boldsymbol{w} = \lambda\boldsymbol{C}\boldsymbol{w}
\end{equation} 
Using $\boldsymbol{M}_{ii}\boldsymbol{w} = \boldsymbol{M}_{,i} \boldsymbol{M}_{,i}^{t} \boldsymbol{w}$
and as $\boldsymbol{M}_{,i}^{t} \boldsymbol{w}$ is a scalar, we have
\begin{equation}
\boldsymbol{C} \boldsymbol{w} \propto \boldsymbol{M}_{,i} \,.
\end{equation}
so that the weights are (up to normalizations) 
\begin{equation}
\boldsymbol{w} = \boldsymbol{C}^{-1} \boldsymbol{M}_{,i} \,.
\end{equation}


For these weights, the Fisher matrix is
\begin{equation}
  \mathcal{F}_{ii} =
  \boldsymbol{M}_{,i}^t\boldsymbol{C}^{-1}\boldsymbol{M}_{,i}
\end{equation}
while the error on $\theta_i$ is
\begin{equation}
  \Delta \theta_i =
  \left(\boldsymbol{M}_{,i}^t\boldsymbol{C}^{-1}\boldsymbol{M}_{,i}\right)^{-1/2} .
\end{equation}
It is worth noting that the Fisher matrix is identical to what we would have without
any compression. Our compression is therefore theoretically lossless, and has extracted
all the information in the original data.

\subsection{A Toy Model}
\label{sec:toy}

In order to build intuition for the results of the previous section, we first apply this to a toy model, which
captures the essential aspects of the full BAO problem. We assume that the correlation function at a separation $x$
and redshift $z$ is a Gaussian 
\begin{equation}
\xi(x,z) = \exp(-x^2).
\end{equation}
We also assume that choosing an incorrect cosmology (here parameterized simply by $\alpha_1$) changes the separation 
$x \rightarrow x - \alpha_1 z$, making the correlation redshift dependent 
\begin{equation}
\xi(x) \rightarrow \xi\left(x-\alpha_1 z\right)
\end{equation}
Our goal is therefore to estimate $\alpha_1$ from observations of $\xi$.

There are three cases we consider. The first (``Full'') is where we imagine
fitting to the correlation function in both $x$ and $z$; this captures all 
the information and should yield the most precise measurements. This would be 
equivalent to splitting the galaxy sample into infinitesimal redshift bins
and fitting a full model to it, which is not possible for the real correlation 
function for the reasons described in Sec.~\ref{sec:intro}. The second (``Unweighted'')
is to integrate the two dimensional correlation function $\xi(x,z)$ along
the $z$ axis to collapse it down to a one dimensional function $\xi_{\rm un}(x) \equiv \int dz \,\xi(x,z)$.

The third case (``Weighted'') is to construct the weighted combination
$\xi_{\rm w}(x) \equiv \int dz \, w(z) \xi(x,z)$ where $w(z)$ are the weights computed in the
previous section. We note that we only combine the data in the $z$ direction here, even
though the formalism of the previous section would have allowed us to collapse in both $x$
and $z$. We do this because the shape of the true galaxy correlation function is uncertain
(hence the marginalization over nuisance shape parameters). Furthermore, as written, the 
weights do not have any $x$ dependence. While not generic, this is explicitly true for our toy
model (as described below). We also construct the weights for the actual correlation function (described in the
next section) to have this property. 

\begin{figure}
	\includegraphics[width=0.5\textwidth]{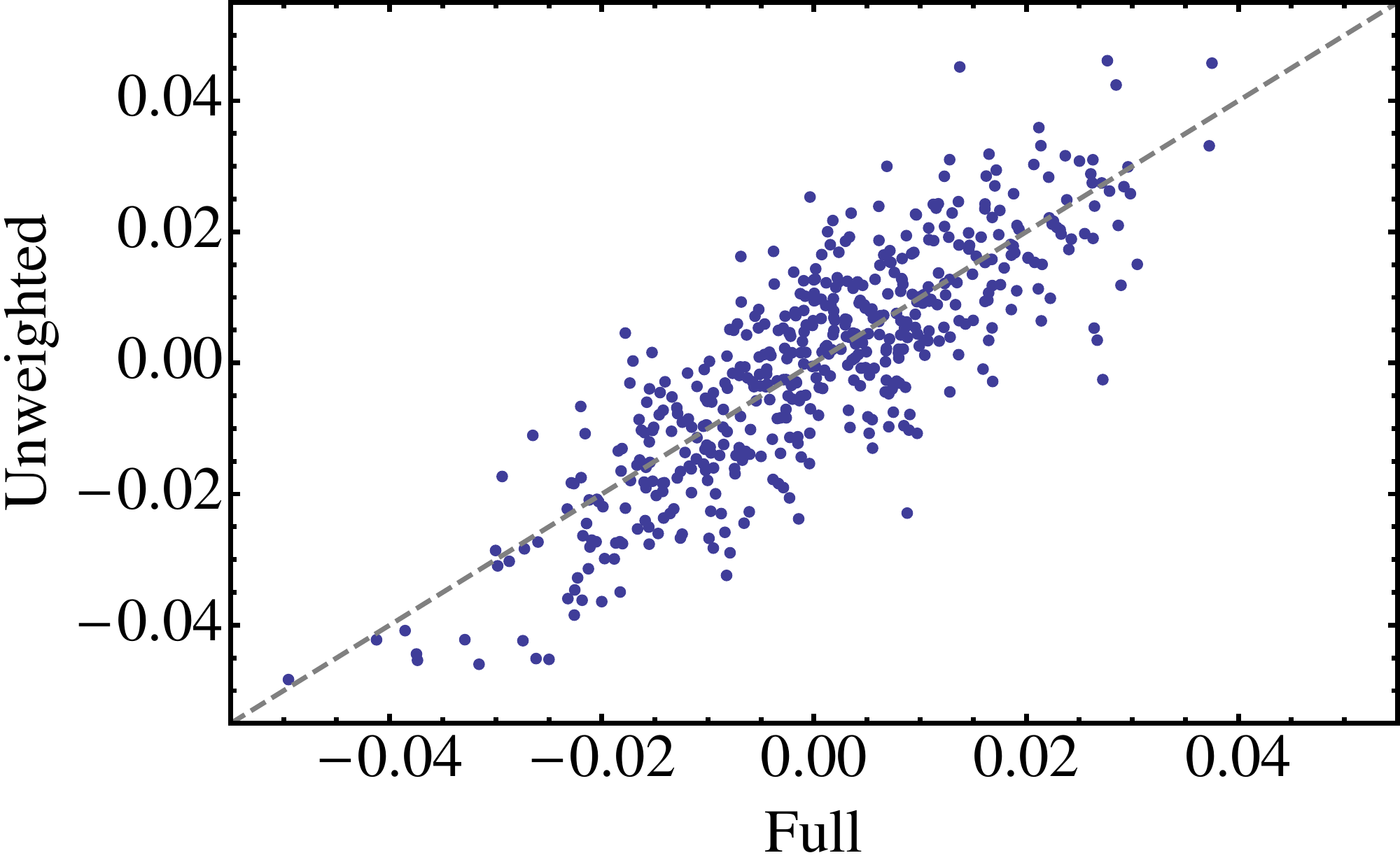}
	\includegraphics[width=0.5\textwidth]{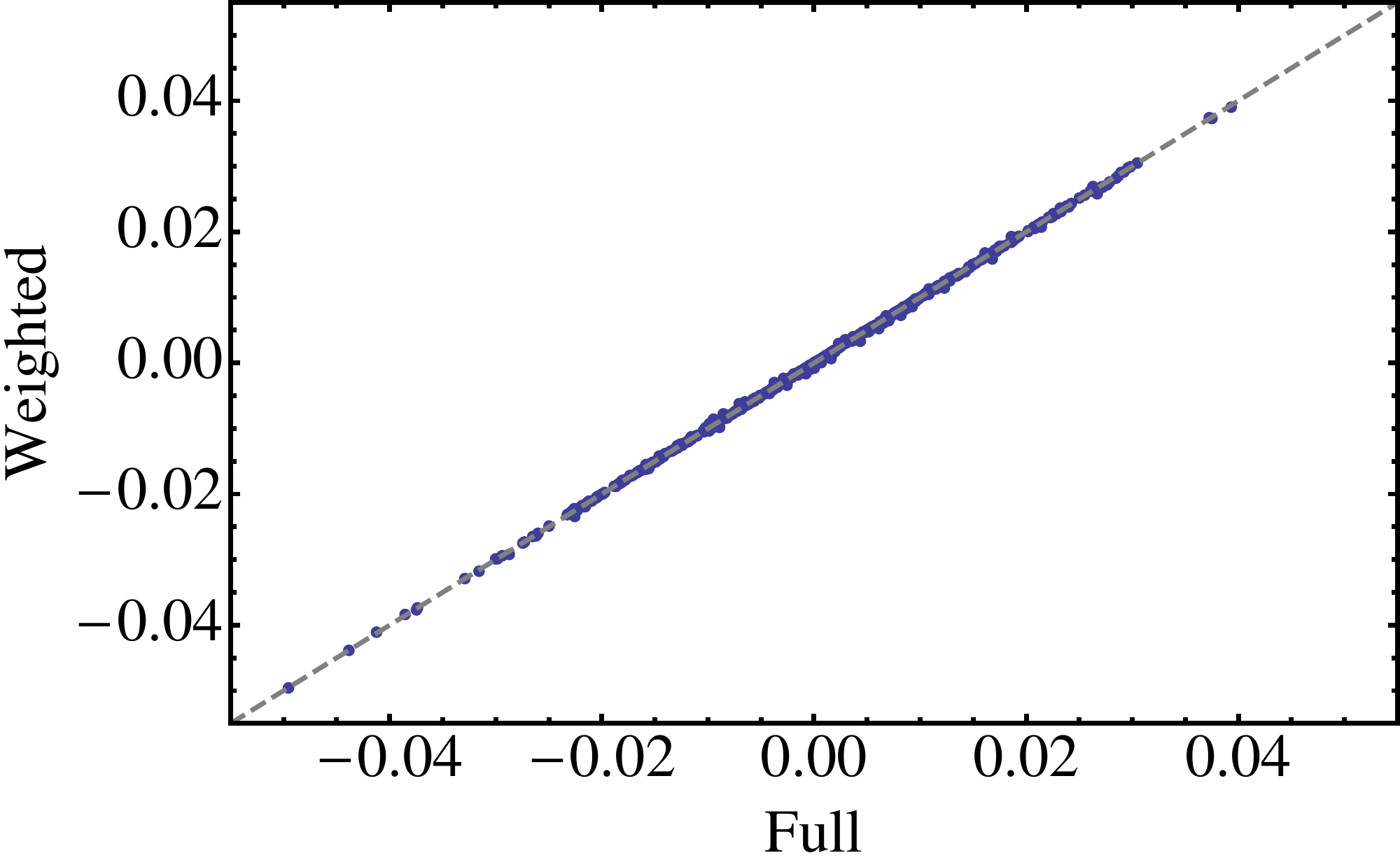}
\caption{A demonstration of the efficiency of using optimal weights to compress information. We 
consider a toy model $\xi(r,z) = \exp(-(x - \alpha_1 z)^2)$, where $\alpha_1$ is the parameter to be fit.
We consider three cases : (1) [x-axis] ``Full'' - where we fit the two-dimensional function
$\xi(x,z)$, (2) [y-axis, top plot] ``Unweighted'' - where we fit $\int dz\,\xi(x,z)$, and (3)
[y-axis, bottom plot] ``Weighted'' - where we fit $\int dz\, w(z) \xi(x,z)$ with $w(z)$ being the
optimal weight. The figure summarizes the results of 500 simulations with the true
value of $\alpha_1$ equal to 0 (see the text for more details).
Both the unweighted and weighted approaches yield unbiased estimates. The weighted estimator is 
almost perfectly correlated with the full case, demonstrating negligible loss of information when
compressing down to a single mode.}
\label{fig:toyfits}
\end{figure}

We can now derive the optimal redshift weights for $\alpha_1$. For simplicity, we will assume 
that the measurements of $\xi(x,z)$ have constant, uncorrelated errors. In this case, 
the weights are simply proportional to $d \xi_{x,z}/d \alpha_1$, giving
\begin{equation}
w(z) = -2 z (x-\alpha_1 z) \exp\left[-(x-\alpha_1 z)^2\right]  \,.
\end{equation}
The above derivative does still depend on $\alpha_1$. We make the usual approximation that
we are making small perturbations to our fiducial cosmology and set $\alpha_1=0$. If this 
were not true (i.e. the measured values of $\alpha_1$ were significantly different from 0),
one would need to iterate using an improved estimate of the true cosmology. Substituting
into the above equation, and using the fact that the normalization of the weights is
arbitrary, we find the simple relation
\begin{equation}
w(z) = z\,.
\end{equation}
This weighting makes physical sense; one wants to downweight measurements close to $z=0$, 
where the correlation function has little sensitivity to $\alpha_1$, and upweight the measurements
as $z$ increases, where the sensitivity is greater. The optimal weight is nothing but a 
matched filter, tracing the redshift dependence of the signal.

In order to quantitatively explore the impact of these weights, we simulate this toy 
model on an $81 \times 11$ grid in $x$ and $z$, with $x$ in $[-4,4]$ and $z$ in $[0,1]$, 
and add a Gaussian random error of $\sigma=0.1$ to each measurement of $\xi$. Note that 
the particular choice of the size of this error does not change any of the conclusions 
below. We verify this by repeating our experiments with $\sigma=1$ and find identical 
behaviour (except for the fact that $\alpha_1$ is more poorly measured overall). We simply 
sum along the $z$ direction weighting each measurement by $1$ and $z$ to obtain the 
corresponding unweighted and weighted simulations respectively. Since we assume the 
errors to be uncorrelated, the variances for the unweighted and weighted simulations are simply
$11 \sigma^2$ and $\sigma^2 \sum_{i} z_i^2$ (where the $z_i$ are the redshifts at which the model is evaluated).
We then estimate $\alpha_1$ by numerically minimizing $\chi^2$, which is the maximum likelihood
estimator in this case.

The results for 500 such simulations are summarized in Fig.~\ref{fig:toyfits}. Each case yields
an unbiased estimate for $\alpha_1$, but the unweighted estimator introduces an additional scatter to
the full case. By contrast, the weighted estimator is almost perfectly correlated with the full 
case, compressing the data to one dimension with virtually no information loss. This highlights the usefulness and
efficiency of using an optimal weighting scheme - the dimensionality of the data can be dramatically
reduced with no loss of information.

\section{Redshift Weighting for BAO}
\label{sec:bao}

\subsection{Preliminaries}
\label{sec:baoprelim}

Motivated by the results of the previous section, we turn to the construction of 
similar weights for BAO measurements. We lay out some basic
preliminaries here, with the derivation in the next section. We will assume that the distance-redshift 
relation is parametrized as in Sec.~\ref{sec:dz}.

In order to construct a redshift dependent model for the correlation function, we
work in the plane parallel approximation and parametrize the effects of an 
incorrect choice of a distance-redshift relationship by an isotropic dilation
($\alpha(z)$) and an anisotropic warping ($\epsilon(z)$) parameter \citep{Padmanabhan08Ani}.
\footnote{Note that an analogous derivation could be carried out for a stretching
in the perpendicular ($\alpha_\perp$) and parallel ($\alpha_\parallel$) directions
as in eg. \cite{Kazin12}.} However, unlike previous analyses which have used this 
parametrization, we will explicitly assume that these are redshift dependent 
(although we will often suppress this dependence for notational brevity).
These distortion parameters can be related to stretching in the perpendicular 
and parallel directions by
\begin{align}
r_\parallel &=\alpha(1+\epsilon)^2 r_\parallel^\text{f}\\
r_\perp &=\alpha(1+\epsilon)^{-1}r_\perp^\text{f}
\end{align}
and the inverse relations by
\begin{align}
\alpha&=\left[\left(\frac{r_\parallel}{r_\parallel^\text{f}}\right)\left(\frac{r_\perp}{r_\perp^\text{f}}\right)^2\right]^{1/3}\\
\epsilon&=\left(\frac{r_\parallel}{r_\parallel^\text{f}}\frac{r_\perp^\text{f}}{r_\perp}\right)^{1/3}-1 \,.
\end{align}
In the plane parallel limit, we have 
\begin{align}
\frac{r_\parallel}{r_\parallel^\text{f}}&=\frac{H_f(z)}{H(z)} \\
\frac{r_\perp}{r_\perp^\text{f}}&=\frac{\chi(z)}{\chi_f(z)}
\end{align}
giving us
\begin{align}
\alpha(z)&=\left[\frac{H_f(z)\chi^2(z)}{H(z)\chi_f^2(z)}\right]^{1/3}\\
\epsilon(z)&=\left[\frac{H_f(z)\chi_f(z)}{H(z)\chi(z)}\right]^{1/3}-1 \,.
\end{align}

Using the definitions above, and the results from Sec.~\ref{sec:dz}, we
now relate $\alpha(z)$ and $\epsilon(z)$ to $\alpha_0$, $\alpha_1$ and $\alpha_2$. We get
\begin{align}
	\alpha &=\left[\alpha_0\left(1+\alpha_1+\left(2\alpha_1+\alpha_2\right)x+\frac{3}{2}\alpha_2x^2\right)\right]^{1/3}\\
&\cdot \left[\alpha_0 \left(1 + \alpha_1 x + \frac{1}{2}\alpha_2 x^2 \right)\right]^{2/3}\\
&\approx \alpha_0\left[1 + \frac{1}{3}\alpha_1 + \frac{1}{3}(4\alpha_1+\alpha_2)x + \frac{5}{6}\alpha_2 x^2 \right]
\end{align}
and
\begin{align}
	1+\epsilon &= \left[\alpha_0\left(1+\alpha_1+\left(2\alpha_1+\alpha_2\right)x+\frac{3}{2}\alpha_2x^2\right)\right]^{1/3} \\
&\cdot \left[\alpha_0 \left(1 + \alpha_1 x + \frac{1}{2}\alpha_2 x^2 \right)\right]^{-1/3}\\
&\approx 1 + \frac{1}{3}\alpha_1 + \frac{1}{3}(\alpha_1+\alpha_2)x + \frac{1}{3}\alpha_2 x^2
\end{align}
where the approximations in the second line are obtained by only working to linear
order in $\alpha_1$ and $\alpha_2$. We will also require the following partial derivatives
(all evaluated at the fiducial value of $\alpha_0=1$, $\alpha_1=\alpha_2=0$ and denoted by subscript $0$)
\begin{align}
	\left.\frac{\partial\alpha}{\partial \alpha_0}\right|_0 &=& 1 & &
\left.\frac{\partial \epsilon}{\partial \alpha_0}\right|_0 &=& 0 \\
   \left.\frac{\partial\alpha}{\partial \alpha_1}\right|_0 &=& \frac{1}{3}(1+4x) & &
\left.\frac{\partial \epsilon}{\partial \alpha_1}\right|_0 &=& \frac{1}{3}(1+x) \\
   \left.\frac{\partial\alpha}{\partial \alpha_2}\right|_0 &=& \frac{1}{6}(2x+ 5x^2) & &
\left.\frac{\partial \epsilon}{\partial \alpha_2}\right|_0 &=& \frac{1}{3}(x+x^2)
\end{align}

\subsection{Derivation}
\label{sec:baoderive}

As was laid out in Sec.~\ref{sec:optderive}, we compress the correlation multipoles
$\xi_{\ell}(r, z)$ along the $z$ direction. The optimal weights are proportional to 
$\boldsymbol{C}\left[\xi_{\ell}(r,z_1), \xi_{\ell}(r,z_2)\right]^{-1} \xi_{\ell,i}$ where
$i$ runs over the parameters $\alpha_0$, $\alpha_1$ and $\alpha_2$. Note that we do not consider 
mixing in the $r$ or $\ell$ directions, to avoid building in sensitivity to the 
model of the correlation function.

To compute $\xi_{\ell,i}$, we need the dependence of the correlation function 
on $\alpha$ and $\epsilon$ \citep{Padmanabhan08Ani,Xu2013} :
\begin{align}
\xi_0(r) &\rightarrow  \xi_0(r)+(\alpha-1)\frac{d\xi_0(r)}{d\log r}\\
\xi_2(r) &\rightarrow 2\epsilon\frac{d\xi_0(\alpha r)}{d\log r}\\
&=2\epsilon\left[\frac{d\xi_0(r)}{d\log r}+(\alpha-1)\frac{d^2\xi_0(r)}{d(\log r)^2}\right]
\end{align}
Note that in the above, we set the intrinsic quadrupole correlation function $\xi_2$ to 
zero. We make this approximation motivated by the fact that we are interested in distance
measurements from the BAO feature, which is very weak in the quadrupole. The derivatives 
of the correlation function are then simply given by 
\begin{align}
	\xi_{0,i} &= \alpha_{,i} \\
	\xi_{2,i} &= \epsilon_{,i}
\end{align}
where we have dropped all non-redshift dependent proportionality constants. We also note that 
the derivatives are all evaluated at the fiducial values of $\alpha=1$ and $\epsilon=0$.

Turning to the inverse covariance matrix, we note that the variance of a correlation function 
bin from a
slice at redshift $z$ is schematically
given by 
\begin{equation}
	C \sim \left(P + \frac{1}{\bar{n}} \right)^2 \frac{1}{dV}
\end{equation}
where 
\begin{equation}
	dV = \frac{\chi_f^2(z)}{H_f(z)}dzd\Omega
\end{equation}
is the volume of slice being considered, $\bar{n}$ is the (redshift dependent)
number density and $P$ is a measure of the power on scales corresponding to
the bin in question.
Again, since we're focusing on the BAO feature, we will ignore the scale
dependence in $P$ and simply treat it as a single number
determined\footnote{One could also treat this as an adjustable parameter
which is empirically determined.} by the approximate scale corresponding
to the BAO feature $(k \sim 0.1 h \rm{Mpc}^{-1})$.
We make the further simplifying assumption that the different redshift
slices are independent, making the covariance matrix diagonal and trivial
to invert: 
\begin{equation}
	d{\cal W}(z) \equiv C^{-1} = \left( \frac{\bar{n}}{ \bar{n} P + 1} \right)^2 dV
\end{equation}
This assumption simply means the weights are not strictly optimal, but will not bias the results.
We observe this weighting is nothing but the commonly used FKP weights \citep{FKP} in galaxy surveys, 
a correspondence we make more explicit in Sec.~\ref{sec:implement}. 
While the normalization of the weights is arbitrary, it is convenient to normalize the weights by the 
integral of the inverse covariance matrix over the survey
\begin{equation}
N = \int d{\cal W} \,.
\end{equation}

We can now put together the different pieces to get our final results. We define the weighted
correlation functions by $\xi_{w, \ell, i}$ :
\begin{equation}
\xi_{w, \ell, i} \equiv \frac{1}{N} \int \, d{\cal W}(z) \, w_{\ell,i}(z) \xi_{\ell}(z)
\label{eq:xidefn}
\end{equation}
where, as before, $i$ runs over the distance-redshift parameters. The $w_{\ell, i}(z)$ are 
given by 
\begin{align}
w_{0,\alpha_0} &=& 1 & &
w_{2,\alpha_0} &=& 0 \\
w_{0,\alpha_1} &=& \frac{1}{3}(1+4x) & &
w_{2,\alpha_1} &=& \frac{1}{3}(1+x) \\
w_{0,\alpha_2} &=& \frac{1}{6}(2x+ 5x^2) & &
w_{2,\alpha_2} &=& \frac{1}{3}(x+x^2)
\end{align}

\begin{figure}
\includegraphics[width=0.5\textwidth]{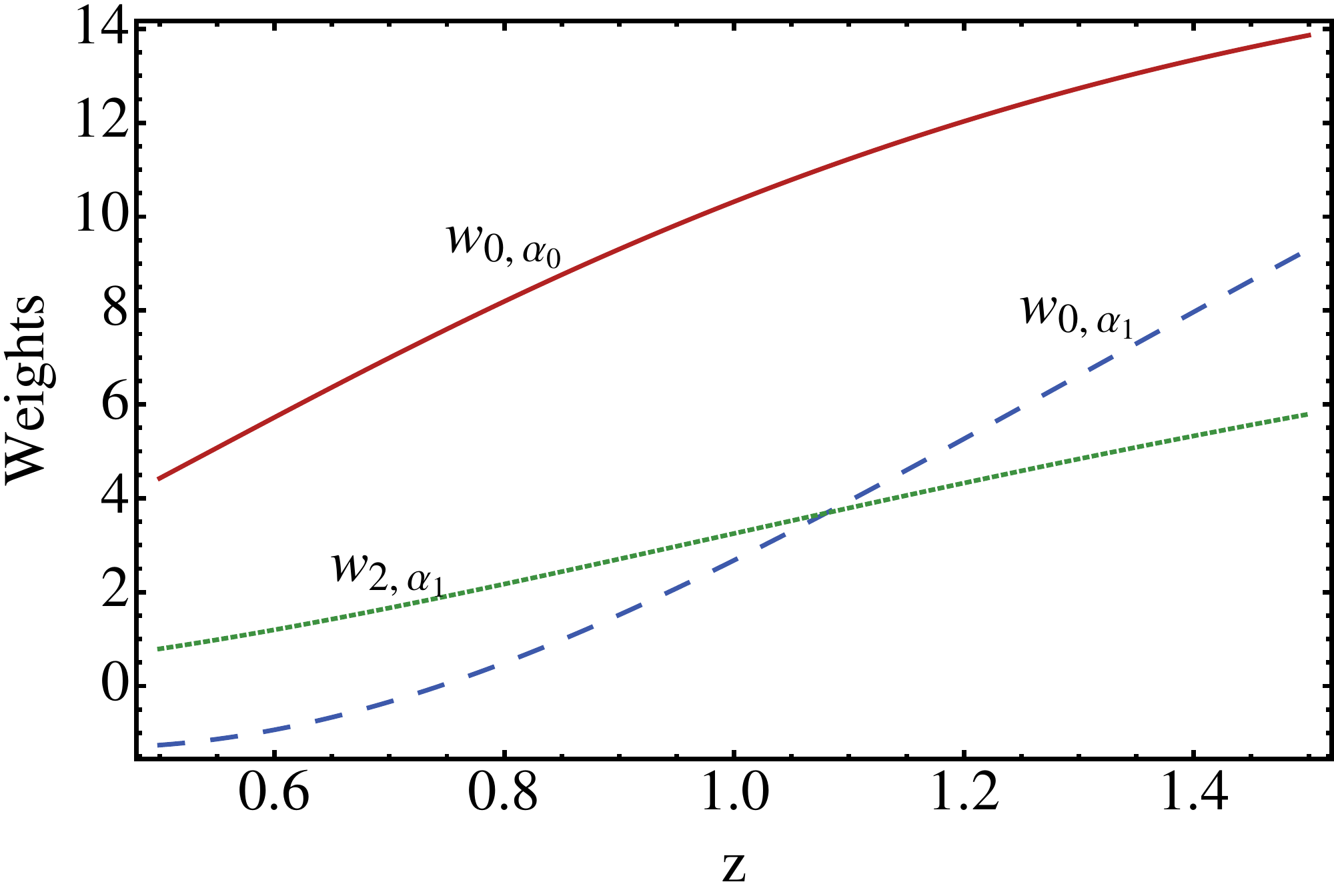}
\includegraphics[width=0.5\textwidth]{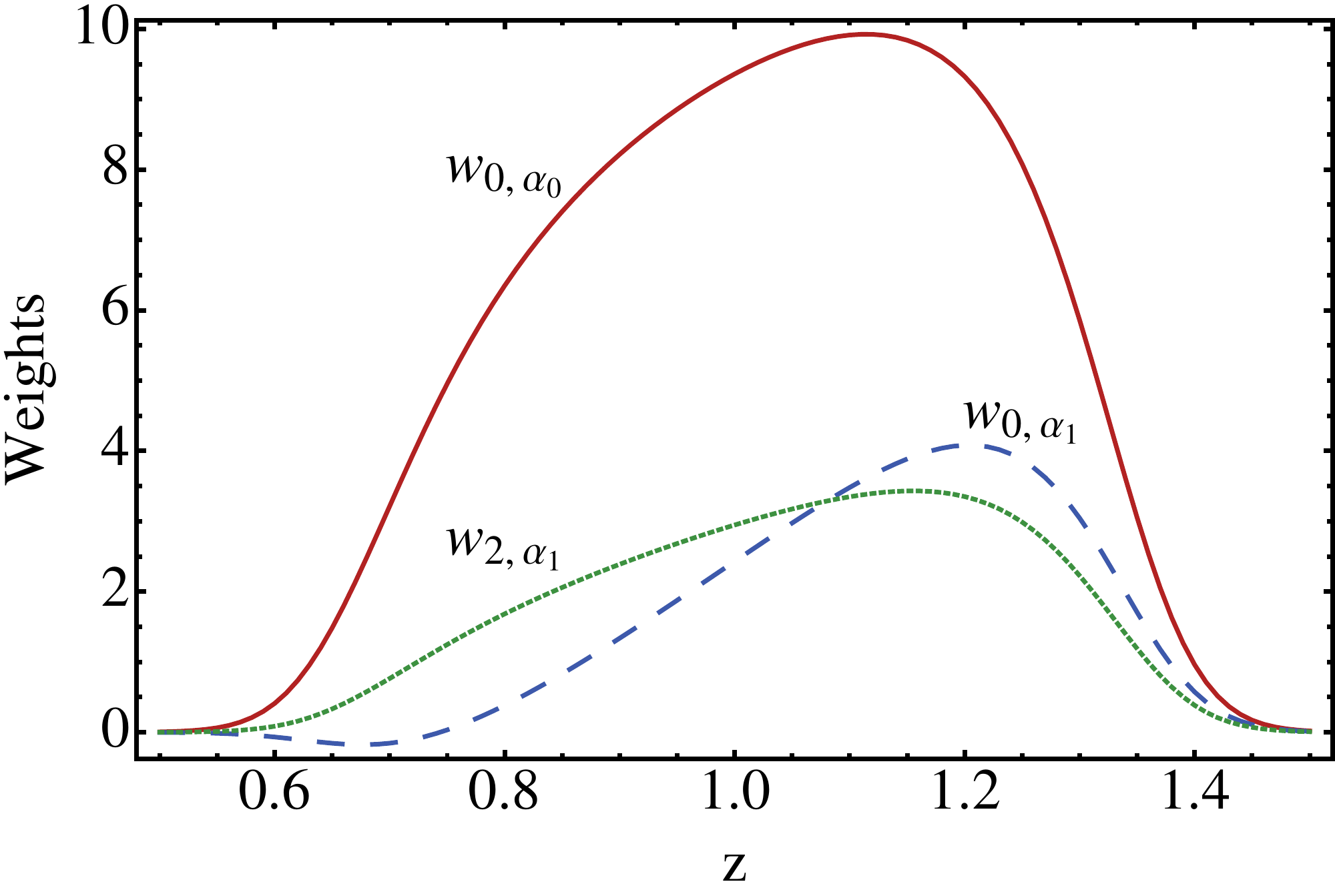}
\caption{ Overall redshift weights for $\alpha_0$ and $\alpha_1$ in monopoles and
	quadrupoles of the two-point correlation function. 
	We plot $w_{\ell,i} d\mathcal{W}$ throughout, although for brevity, 
	the curves are simply labeled by $w_{\ell,i}$.
The top panel assumes a
	constant $\bar{n}$ survey. The solid line 
is for $w_{0,\alpha_0}$ and is just the inverse variance weights, the dashed is $w_{0, \alpha_1}$ 
and the dotted is $w_{2,\alpha_1}$. The bottom panel assumes a gaussian $\bar{n}$
centered around $z=1$. The normalization of the weights is arbitrary. We have
boosted the amplitudes of the weights for better visualization. }
\label{fig:examples}
\end{figure}

Eq.~\ref{eq:xidefn} makes explicit that the overall redshift weights are the
product of $w_{\ell, i}(z)$ and $d\mathcal{W}(z)$. We show two examples in
Fig.~\ref{fig:examples}. The top panel plots the weights for a constant
$\bar{n}$ survey. The bottom panel is the same plot assuming a gaussian
$\bar{n}$ centered around $z=1$. In a constant $\bar{n}$ survey, the weights
for $\alpha_0$ are simply the inverse variance weights. The upward sloping weights
curve suggest an upweight in the high redshift region of the survey. This is
consistent with the intuition that slices at higher redshifts contain more
volume and more galaxies, and thus smaller error bars, than those of lower
redshifts.

\subsection{Interpretation and Examples}
\label{sec:interpret}

In order to develop some intuition for these weighted correlation functions, we consider the correlation
function measured perpendicular to the line of sight; the case parallel to the line of sight is analogous.
This case is simpler than the multipoles presented above, since it doesn't mix the angular diameter
distance and Hubble parameter measurements. In particular, we have 
\begin{equation}
	\xi_{\perp}(r) \rightarrow \xi_{\perp}(\alpha_{\perp} r) \sim \xi_{\perp}(r) + (\alpha_{\perp}-1) \frac{d \xi_{\perp}}{d \log r}
\end{equation}
where 
\begin{equation}
	\alpha_{\perp} \equiv \frac{\chi(z)}{\chi_f(z)} = \alpha_0 \left(1 + \alpha_1 x + \frac{1}{2} \alpha_2 x^2 \right) \,.
\end{equation}
The corresponding weights are then
\begin{align}
w_{\alpha_0} &= 1 \\
w_{\alpha_1} &= x \\
w_{\alpha_2} &= x^2/2 \,.
\end{align}
The weighted correlation functions can be schematically written as 
\begin{equation}
	\xi_w = A \xi_{\perp} + B \frac{d \xi_{\perp}}{d \log r}
\end{equation}
where $A$ is just a normalization constant (independent of the values of $\alpha_0$, $\alpha_1$ and $\alpha_2$) and 
$B$ are integrals depending on the particular weights and $\alpha_0$, $\alpha_1$ and $\alpha_2$. The derivative
term in the above equation is equivalent to a shift of the BAO feature. The different redshift weights result in
different shifts for a given set of distance parameters. From these measured shifts, one can then infer the 
underlying parameters.

To make this example more concrete, we consider a redshift range defined by $-1 \le x \le 1$ and assume the 
inverse variance weights are simply constant. The integrals are then trivial and are given by:
\begin{align}
	B_{\alpha_0} &\propto 2 \alpha_0 + \frac{\alpha_0 \alpha_2}{3} \\
	B_{\alpha_1} &\propto \frac{2}{3}\alpha_0 \alpha_1 \\
	B_{\alpha_2} &\propto \frac{1}{3} \alpha_0 + \frac{1}{10} \alpha_0 \alpha_2
\end{align}
where we have focused just on the shifts in the BAO feature. From the above, it is clear that given the
measurements of three shifts, one can invert these to determine the $\alpha_0$, $\alpha_1$ and $\alpha_2$.

\subsection{Alternate Parametrizations}

\begin{figure}
\includegraphics[width=0.5\textwidth]{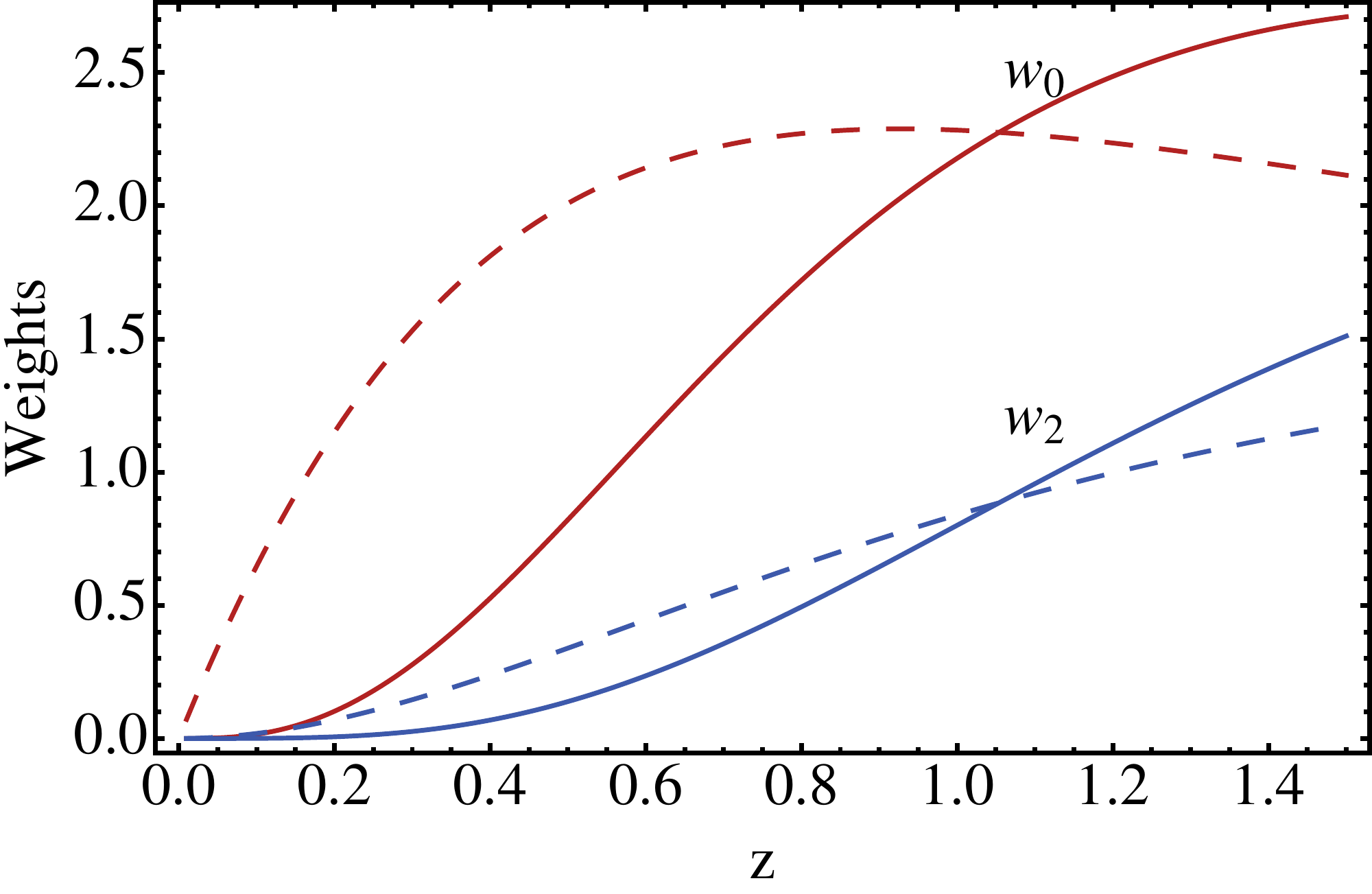}
\caption{The optimal weights to constrain the dark energy equation of state $w$, for the monopole ($w_0$ [red]) and
the quadrupole ($w_2$ [blue]). We assume a survey with a constant number density. 
As before, the normalizations here are arbitrarily chosen for plotting convenience.
The solid lines include the $d\mathcal{W}$ volume factor, while the dashed lines do not.
The figure shows the relative importance of different redshifts to measuring $w$. At low redshifts, the weights are
suppressed both due to the small volume, as well as the fact that all distance redshift relations tend to their Hubble
law form $\chi = (c/H_0) z$ independent of the equation of state. The impact of the volume factor with increasing redshift is also
clearly evident.
We note that this is an illustrative example; a 
complete treatment would include parameters beyond $w$.}
\label{fig:weight_w}
\end{figure}

Although we recommend our parametrization of the distance-redshift relation, the idea of optimal redshift weights is more
general. To illustrate this, we consider a standard $w$CDM cosmology, with the Hubble parameter (and therefore, the
distance-redshift relation) parametrized by a constant equation of state $w$ for dark energy :
\begin{equation}
	H^2(z) = H_0^2 \left[ \Omega_M (1+z)^3 + (1-\Omega_M) (1+z)^{3(1+w)} \right] \,.
\end{equation}
Since this is illustrative, we hold $H_0$ and $\Omega_M$ fixed; a full treatment would vary multiple parameters.

Proceeding as before, the weights will depend on derivatives of the monopole and quadrupole with $w$, which in turn, 
are proportional to derivatives of $\alpha(z)$ and $\epsilon(z)$ with $w$. Fig.~\ref{fig:weight_w} plots these weights
for a survey with a constant number density of objects. 
The figure highlights the relative importance of different redshifts to measuring $w$. At low redshifts, the weights are
suppressed both due to the small volume, as well as the fact that all distance redshift relations tend to their Hubble
law form $\chi = (c/H_0) z$ independent of the equation of state. 
The impact of the volume factor with increasing redshift is also clearly evident, upweighting the higher redshift regions
even when the raw sensitivity to $w$ is flattening out (or decreasing).

\subsection{Implementation Notes}
\label{sec:implement}

We briefly discuss a possible implementation of this weighting scheme. We do not claim 
that this is a final implementation and we defer tests with mock catalogues to future work. Our goal 
here is to provide a proof-of-principle that such an implementation is both possible and straightforward.

We start with Eq.~\ref{eq:xidefn}, and write the integral as a sum over infinitesimal redshift slices 
indexed by $z$. We further
write the correlation function as a combination of data and random pairs, using the \cite{LS} estimator :
\begin{equation}
\xi_{w, \ell, i} = \frac{1}{N} \sum_{z} \, \Delta{\cal W}_{z} \, w_{\ell,i,z} 
\left[ \frac{DD_z - 2 DR_z + RR_z}{RR_z} \right] \,
\end{equation}
where $DD_z$ are the data-data pairs in the particular $z$ bin etc. Since the weights 
$\Delta {\cal W}$ and $w_{\ell,i}$ only depend on the redshift slice, the estimator 
could have been written as a weighted pair sum if not for the $RR_z$ term in the denominator. 
One solution is to construct a model for $RR$ as proposed in \cite{PadmanabhanOmega}. We 
consider a different approximation below.

Consider the ratio $\Delta {\cal W}_z/RR_z$. The random-random pairs $RR_z$ can be written
as 
\begin{equation}
RR_z = \bar{n}^2 \Phi \Delta V
\end{equation} 
where $\Phi$ encodes the effect of the survey geometry on the pair counts. The ratio is then
\begin{equation}
\label{eq:dW/RR}
\frac{\Delta {\cal W}_z}{RR_z} = \left( \frac{1}{1 + \bar{n} P} \right)^2 \frac{1}{\Phi} \,.
\end{equation}
In the limit that $\Phi$ is independent of redshift, it can be factored out of the sum. In 
this case, the terms in the numerator are a simple pair sum with every galaxy/random weighted
by $w_\text{FKP}(z) = (1+\bar{n} P)^{-1}$ and every pair by $w_{\ell, i}(z)$. 
Namely, we compute the weighted pair sums as
\begin{align}
\widetilde{DD} &= \sum_z w_{\ell,i,z} w_\text{FKP}^2(z) DD_z \\
\widetilde{DR} &= \sum_z w_{\ell,i,z} w_\text{FKP}^2(z) DR_z \\
\widetilde{RR} &= \sum_z w_{\ell,i,z} w_\text{FKP}^2(z) RR_z 
\end{align}
Applying a similar argument to the $N \Phi$ term in the denominator by using Eq.~\ref{eq:dW/RR},
\begin{equation}
N\Phi = \Phi \sum_z \Delta {\cal W}_z = \sum_z w_\text{FKP}^2(z)  RR_z \,.
\end{equation}
We see that this is nothing but the usual random-random
pairs $RR$, with every random weighted by $(1+\bar{n} P)^{-1}$. 

The weighted correlation function estimator can now be written succinctly as
\begin{equation}
\label{eq:weighted_xi}
\xi_{\ell,i} = \frac{\widetilde{DD} - 2\widetilde{DR} + \widetilde{RR}}{RR}
\end{equation} 

Our proposed algorithm is :
\begin{enumerate}
\item Weight every galaxy/random by $w_\text{FKP} = (1 + \bar{n} P)^{-1}$. 
\item Compute $\widetilde{DD}$, $\widetilde{DR}$ and $\widetilde{RR}$ as a weighted pair sum, including $w_{\ell, i}(z)$ as an
additional pair
weight. Since the change in redshift is small for the scales of interest, one could simply 
use eg. the mean redshift of the pair in computing $w_{\ell,i}$.
\item Compute $RR$ as a weighted pair sum including $w_\text{FKP}$, but not the $w_{\ell,i}(z)$ weight.
\item Compute the correlation function estimator according to Eq.~\ref{eq:weighted_xi}.
Note that in the absence of the $w_{\ell,i}$, $\widetilde{DD} \to DD$, $\widetilde{DR} \to DR$, $\widetilde{RR} \to RR$. We recover the usual algorithm
for computing correlation functions.
\item A final practical simplification : since the weights are simply linear
	combinations of $1$, $x$ and $x^2$, it is more efficient to compute correlation
	functions weighted by them instead of the original weights. 
\end{enumerate}

As an aside, the above discussion makes it clear how the usual $(1+\bar{n} P)^{-1}$
weights used in correlation function analyses are an approximation (under certain
assumptions) to the optimal $C^{-1}$ weighting \citep{Hamilton93}.

\section{Discussion}
\label{sec:discuss}

Large redshift surveys capable of measuring the baryon acoustic oscillation
signal have proven to be an effective way of measuring the distance-redshift
relation in cosmology.
Future BAO surveys will survey very large volumes, covering wide ranges in
redshift.
The question arises how best to analyze and interpret the information from
such a wide range of $z$.
One approach is to split the sample into a number of smaller redshift shells,
with the BAO signal in each shell providing a measurement of the distance to
a fiducial redshift within the shell.  The drawback of such a procedure is
that the BAO signal in each shell is of lower signal-to-noise, and this
makes the analysis more sensitive to the tails of the distribution.
Since variations in the distance redshift relation are very smooth in most
cosmologies, we advocate a different procedure.  Specifically we suggest
compressing the information in the redshift direction into a small number
of `weighted correlation functions'.
We find that a small number of redshift weights preserves almost all of the
information, with little dilution in signal-to-noise per measurement.

We demonstrate that the distance-redshift relation, relative to a fiducial
model, can be approximated by a quadratic function
to sub per cent accuracy over a broad range of cosmologies and redshifts. 
The coefficients can be straightforwardly related to the distance-redshift
relation and its first two derivatives at a reference redshift $z_0$. 
This reference redshift is now an explicit choice and resolves ambiguities
in the interpretation of the ``effective'' redshift of a BAO measurement.
For this particular choice of parameters, we then construct weights 
to optimally constrain their values from the data. We explicity demonstrate 
the process with toy models, and outline a possible implementation of this 
approach.

Our parametrization of the distance-redshift relation is not unique. For 
instance, one could choose a parametrization based on the dark energy 
equation of state. Different choices would result in different weights, 
although their construction would be identical to what we present here.
These differences are however superficial. The key message of this paper
remains the same - that the parameters of the expansion history may
be constrained by a relatively small number of modes.

This paper has made the simplifying assumption that the true 
correlation function does not have any redshift dependence. This is
clearly not true, especially for samples selected using different
techniques and for wide redshift ranges. This problem is not 
unique to this approach though. We expect it can be solved in 
part constructing a model for this redshift evolution from eg. 
small scale clustering measurements, and in part by introducing 
additional nuisance parameters to absorb remaining discrepancies.
The exact details will depend on the particular sample and 
we defer further investigations to future work. We do point out 
that an interesting open question is how well this redshift 
evolution needs to be known to not compromise the fidelity and
precision of future BAO measurements.

This paper lays the framework for a new approach to fitting BAO 
measurements. We plan on continuing to develop this approach in 
future work by testing possible implementations on mock catalogs,
and ultimately applying it to existing surveys.

\section{Acknowledgments}

We thank Chris Blake, Daniel Eisenstein and Will Percival for useful discussions.
FZ, NP and MW acknowledge the Santa Fe Cosmology Workshop where part of
this work was completed. 
This work was supported in part by the National Science Foundation under 
Grant No. PHYS-1066293 and the hospitality of the Aspen Center for Physics.
NP is supported in part by DOE, NASA and a Sloan Research Fellowship.

\bibliographystyle{mn2elong}
\bibliography{paper1}

\appendix

\section{Derivation of the distance-redshift parameters}
Our distance redshift relation in Sec. \ref{sec:dz} can be written as 
\begin{equation}
\label{eq:r(z)}
	\chi(z) = \alpha_0 \left( 1 + \alpha_1 x + \frac{1}{2}\alpha_2 x^2 \right) \chi_f(z) .
\end{equation}
Taking the derivative of both sides with respect to $z$ by using
\begin{equation}
\label{eq: dx}
	\frac{dx}{dz}=\frac{1}{H_f(z)\chi_{f,0}},
\end{equation}
we have 
\begin{align}
\label{eq: first_deriv}
\frac{1}{H(z)} &= \alpha_0 \left[ \left( 1 + \alpha_1 x + \frac{1}{2}\alpha_2 x^2 \right)\frac{1}{H_f(z)}  \right.\nonumber \\ 
			   & \left. + \left(\alpha_1 + \alpha_2 x \right) \frac{\chi_f(z)}{H_f(z)\chi_{f,0}}\right] 
\end{align}
Evaluating at $z = z_0$ ($x$ vanishes), we have 
\begin{equation}
	\frac{1}{H_0} = \alpha_0 \left(\frac{1}{H_{f,0}} + \frac{\alpha_1}{H_{f,0}} \right).
\end{equation}
Solving for $\alpha_1$ yields 
\begin{equation}
\label{eq:b1}
	\alpha_1 = \frac{H_{f,0}\chi_{f,0}}{H_0 \chi_0} - 1.
\end{equation}
Another useful quantity is $H_f(z)/H(z)$, which can also be derived from (\ref{eq: first_deriv}). Together with the definition of $x$, 
\begin{align}
	\frac{H_f(z)}{H(z)} &= \alpha_0\left[ 1+\alpha_1 x+\frac{1}{2}\alpha_2 x^2 + (\alpha_1+\alpha_2 x) (1+x) \right] \\
	&= \alpha_0\left[1+\alpha_1+\left(2\alpha_1+\alpha_2\right)x+\frac{3}{2}\alpha_2x^2\right]
\end{align}
To calculate $\alpha_2$, we take the second derivative of (\ref{eq:r(z)}) and evaluate at $z = z_0$, 
\begin{equation}
	\chi''(z_0) = \alpha_0 \left[\chi''_f(z_0) + 2 \alpha_1 x'_0 + \chi_{f,0}\left(\alpha_1 x''_0 + \alpha_2 x'^2_0 \right)\right]
\end{equation}
where $x' = dx/dz$ and $x'' = d^2 x/ dz^2$. The subscript 0 indicates the derivatives are all evaluated at $z = z_0$.
Solving for $\alpha_2$ by plugging in (\ref{eq: dx}), and using $\chi'(z) = 1/H(z)$, we obtain
\begin{equation}
	\alpha_2 = (1+\alpha_1)\chi_{f,0}\left[H'_{f,0} - \alpha_0(1+\alpha_1)H'_0\right] - 2\alpha_1
\end{equation} 

\end{document}